\documentstyle[prd,aps,tighten,preprint]{revtex}

\begin{document}
\draft
\title{Flavor symmetry breaking in quark magnetic
moments\footnote{Supported
by the Swedish Natural Science Research
Council (NFR), contract F-UR32821-105.}}

\author{Johan Linde and H{\aa}kan Snellman}
\address{Department of Theoretical Physics\\
Royal Institute of Technology\\
S-100 44 STOCKHOLM\\
SWEDEN}

\maketitle

\begin{abstract}
We discuss the magnetic moments of the baryons allowing for flavor symmetry
breaking
in the quark magnetic moments. We show that there is a correlation between
isospin
symmetry breaking
and data for the nucleon spin structure obtained from deep inelastic
scattering. For small
values of the isospin symmetry breaking, of the order of $5\%$, the
magnetic moments
and weak axial-vector form factors alone indicate a value of the
spin polarization $\Delta\Sigma \leq 0.20$.
Larger values of the spin polarization are compatible only with large
isospin
symmetry breaking.
We also calculate weak axial-vector form factors, which are independent
of the
symmetry breakings, from magnetic moment data and find good agreement
with experiment.
\end{abstract}


\narrowtext

\section{Introduction}
The non relativistic quark model (NQM) is a very efficient tool for
correlating
low energy hadron data. For example, calculating the magnetic moments of
the hadrons
in the NQM gives surprisingly good results, indicating that even the
detailed structure
of the hadrons is well described. However, deep inelastic scattering
experiments indicate
that only a fraction of the nucleon spin is carried by the quarks, although
in the NQM
they alone carry the full spin of the hadron. This leads to a more
complex picture of the hadron, being composed of valence quarks,
quark-antiquark pairs
and gluons in some mixture.

At a more refined level of analysis the NQM must therefore be modified. The
most
important such modification concerns the distribution of the spin
polarization
of the baryons on their constituents. This spin structure is reflected e.g.
in the
magnetic moments, in the weak axial-vector form factors and in the deep
inelastic
scattering data.

Let us define the spin projection numbers $n_q^{\uparrow}$ and
$n_q^{\downarrow}$
\begin{equation}
n_q^{\uparrow\downarrow}=3D\int_0^1q^{\uparrow\downarrow}(x)\,dx,
\end{equation}
where $x$ is the fraction of the baryon momentum carried by the quark and
$q^{\uparrow\downarrow}(x)$ is the quark spin distribution function.
 The difference of these is
\begin{equation}
\Delta n_q =3Dn_q^{\uparrow} - n_q^{\downarrow}.
\end{equation}
Similarly we have for
the antiquarks
\begin{equation}
\Delta n_{\bar q} =3Dn_{\bar q}^{\uparrow} - n_{\bar q}^{\downarrow}.
\end{equation}
The spin contribution from a quark to the magnetic moment of a baryon
is given by
$\Delta n_q - \Delta n_{\bar q}$,
whereas the spin contribution from a quark
measured in weak decays and in polarized deep inelastic scattering
experiments is given by
\begin{equation}
\Delta q =3D \Delta n_{q} + \Delta n_{\bar q},
\end{equation}
where $q=3Du,d,s$. The difference in signs comes from the difference under c=
harge
conjugation. As has been discussed by Karl\cite{karl} there is a large class=
 of
models in which one can use the $\Delta q$'s in the formulas for magnetic
moments. The quark magnetic moments are then taken to be effective moments.
This allows a general parameterization of all the data for nucleon spin
structure and will be adopted here.

In the NQM the quarks carry properties, such as charge, mass and magnetic
moment, which are independent of their surroundings. Experimentally this
seems not fully to be the case.
We will therefore  consider a model where the magnetic moments of the quarks
are different in  different isomultiplets, but do not
change within an isomultiplet. This is reasonable if one wants to keep
at least approximately independent quark properties.
This point will be discussed further in Section \ref{sec2}.

Keeping the symmetry for the spin projection numbers, i.e., $\Delta u^N=3D\D=
elta
s^\Xi$
etc.,
the magnetic moments are given by
\begin{mathletters}
\begin{eqnarray}
\mu(p)=3D&&\mu_u^N\Delta u+\mu_d^N\Delta d+\mu_s^N \Delta s,\\
\mu(n)=3D&&\mu_d^N\Delta u+\mu_u^N\Delta d+\mu_s^N \Delta s,\\
\mu(\Sigma^+)=3D&&\mu_u^\Sigma\Delta u+\mu_s^\Sigma\Delta d+\mu_d^\Sigma
\Delta s,\\
\mu(\Sigma^-)=3D&&\mu_d^\Sigma\Delta u+\mu_s^\Sigma\Delta d+\mu_u^\Sigma
\Delta s,\\
\mu(\Xi^0)=3D&&\mu_s^\Xi\Delta u+\mu_u^\Xi\Delta d+\mu_d^\Xi \Delta s,\\
\mu(\Xi^-)=3D&&\mu_s^\Xi\Delta u+\mu_d^\Xi\Delta d+\mu_u^\Xi \Delta s.
\end{eqnarray}
\end{mathletters}

It would in principle be possible to let the $\Delta q$'s also be different
in different isomultiplets. However,
since the baryon magnetic moments
depend on the product of the spin polarizations and the quark magnetic
moments,
it is clearly possible to compensate a change in the spin structure by a
change in
magnitude of the quark magnetic moments. Thus, the above parameterization is
the most general. In fact, the deviation of the $\Delta q$'s from the NQM
values comes from gluonic components in the wave-function, which should be
independent of flavor. This is the reason for keeping the symmetry for
the $\Delta q$'s.
Later on we will discuss weak axial-vector form factors and
see that the above spin polarizations are in fact the same
as the ones in the expressions for these form factors.

We see that it is  not possible to
determine
the quark magnetic moments unless the spin polarizations are known.
In fact, the above expressions can be taken as a definition of the (effectiv=
e)
quark magnetic moments.
The weak axial-vector form factors are dependent only on two spin
polarization
differences $\Delta u -\Delta d$ and $\Delta u -\Delta s$. It is therefore
not enough
to use these form factors as input. One still needs data from
deep inelastic scattering to determine the quark magnetic moments.

Most previous analyses of baryon magnetic moments
\cite{karl,avenarius,sehgal,bartelski,decker}
have been made assuming isospin symmetry in the quark magnetic moments.
=46or a general analysis it is of interest to admit also for isospin symmetr=
y
breaking.
Lattice calculations, for example, indicate a relatively large isospin
symmetry breaking of the order of $10 \%$ \cite{leinweber}. We will
therefore
allow for isospin as well as hypercharge symmetry breaking in this model.
As was mentioned before, the quark magnetic moments, and thus this symmetry
breaking,
will be determined only when data from deep inelastic scattering is
considered. Our
analysis favors rather small values of $\Delta \Sigma$, the parameter that
is a measure of the quark contribution to the nucleon spin, unless
the isospin symmetry breaking is large. For values of $\Delta \Sigma \geq
0.35$ the isospin symmetry breaking is of the
order of $10 \%$. Below we first treat the magnetic moments and their
relations in the model. We
then discuss the weak axial-vector form factors and connect them to the
magnetic moments. At last
we study the nucleon spin polarization as a function of the isospin
symmetry breaking.

\section{Magnetic moments}\label{sec2}

Let us first argue why the quark magnetic moments are different in
different isomultiplets.
If the three quark magnetic moments  were the same
in all baryons we find the sum-rule \cite{franklin}
\begin{equation}
(0.38\pm0.04\ \mu_N)\quad\mu(p)+\mu(\Sigma^-)+\mu(\Xi^0)=3D
\mu(n)+\mu(\Sigma^+)+\mu(\Xi^-)\quad (-0.11\pm0.02\ \mu_N),
\label{summaregel}
\end{equation}
which is definitely violated experimentally. This shows that the $\mu_q$'s
are not constant, as they are in the NQM.

Now, the parameterization of the nine $\mu_q$'s must be such that there
are at most five independent parameters to be able to uniquely solve for
them. On the other hand, we still want to keep the number of degrees of
freedom as large as possible. A symmetric way of parameterizing the
moments is to introduce symmetry breakings $T$ and $U$ that are the same
for all isomultiplets, i.e.
$T=3D\mu_u^B/\mu_d^B$
and $U=3D\mu_s^B/\mu_d^B$, $B=3DN,\Sigma,\Xi$.

The number
of magnetic moment parameters are now five: $\mu_d^N,\mu_d^\Sigma,
\mu_d^\Xi,T,U$. To determine the magnetic moments we also need the three
spin polarizations.
The system of equations now looks as follows:
\begin{mathletters} \label{mu}
\begin{eqnarray}
\mu(p)=3D&&\mu_d^N(T\Delta u+\Delta d+U\Delta s),\\
\mu(n)=3D&&\mu_d^N(\Delta u+T\Delta d+U\Delta s),\\
\mu(\Sigma^+)=3D&&\mu_d^\Sigma(T\Delta u+U\Delta d+\Delta s),\label{musp}\\
\mu(\Sigma^-)=3D&&\mu_d^\Sigma(\Delta u+U\Delta d+T\Delta s),\label{musm}\\
\mu(\Xi^0)=3D&&\mu_d^\Xi(U\Delta u+T\Delta d+\Delta s),\\
\mu(\Xi^-)=3D&&\mu_d^\Xi(U\Delta u+\Delta d+T\Delta s).
\end{eqnarray}
\end{mathletters}
With $T=3D-2$, i.e.\ isospin symmetry,
 this reduces to the equations used by Avenarius \cite{avenarius},
and if we in addition equate the $\mu_d$'s  we get the model used by
Karl \cite{karl}.

 Eliminating the three $\mu_d$-parameters from the
equations (\ref{mu}) we get
\begin{mathletters}\label{egenekv}
\begin{eqnarray}
2U\Delta u+(1+T+C(1-T))\Delta d+(1+T+C(T-1))\Delta s=3D&&0, \label{egen1}\\
(1+T+B(1-T))\Delta u+2U\Delta d+(1+T+B(T-1))\Delta s=3D&&0, \label{egen2}\\
(1+T+A(1-T))\Delta u+(1+T+A(T-1))\Delta d+2U\Delta s=3D&&0,
\end{eqnarray}
\end{mathletters}
where
\begin{eqnarray}
A=3D&&\frac{\mu(p)+\mu(n)}{\mu(p)-\mu(n)},\\
B=3D&&\frac{\mu(\Sigma^+)+\mu(\Sigma^-)}{\mu(\Sigma^+)-\mu(\Sigma^-)},\\
C=3D&&\frac{\mu(\Xi^0)+\mu(\Xi^-)}{\mu(\Xi^0)-\mu(\Xi^-)}.
\end{eqnarray}
This set of equations (\ref{egenekv}) only has a nontrivial solution if
the secular determinant is $=3D0$, which gives
$U=3D-1-T$ or $U=3D(1+T\pm(1-T)\sqrt{AB-AC+BC})/2$. The first root
$U=3D-1-T$ is a false root which gives $\Delta u=3D\Delta d=3D\Delta s$, whi=
ch
implies that all the right hand sides of (\ref{mu}) are equal to zero.
The root $U=3D(1+T-(1-T)\sqrt{AB-AC+BC})/2$ has the value $U<0$
 for all values of $T<-0.13$. Since we expect $T$ to be less than that
and $U>0$, this root can also be discarded.
Therefore we conclude that
\begin{equation}
U=3D\frac{1}{2}(1+T+D(1-T)), \label{uekv}
\end{equation}
where
\begin{equation}
D=3D\sqrt{AB-AC+BC}.
\end{equation}

=46or the magnetic moments we take data from \cite{data} except for
$\mu(\Sigma^+)$, for which
we use the latest world average \cite{average}.
Numerically we get $U=3D(0.89\pm0.01)+(0.11\pm0.01)T$. Thus $U$ cannot
determine $T$ very accurately, whereas already a rough determination of
$T$,
e.g. $T=3D-2$, is sufficient to determine $U$.

To determine all the eight parameters we need two more
independent data. Unfortunately these cannot be taken from
the magnetic moment of $\Lambda$
and the transition magnetic moment $\mu(\Sigma \Lambda)$ since they both
are independent of the symmetry breakings in the model.
In fact they can essentially be expressed in terms of the three
constants $A,B$ and $C$. The argument goes as follows.

Since $\Lambda$ and $\Sigma$ have the same strangeness,
it is reasonable to assume that the quark magnetic moments
in them are the same.
The magnetic moment of $\Lambda$ and the transition magnetic
moment $\mu(\Sigma \Lambda)$ are then given by
\begin{equation}
\mu(\Lambda)=3D\frac{1}{6}(\Delta u+4\Delta d+\Delta s)(\mu_u^\Sigma+
\mu_d^\Sigma)
+\frac{1}{3}(2\Delta u-\Delta d+2\Delta s)\mu_s^\Sigma, \label{muL}
\end{equation}
\begin{equation}
\mu(\Sigma \Lambda)=3D-\frac{1}{2\sqrt{3}}(\Delta u-2\Delta d+\Delta s)
(\mu_u^\Sigma-\mu_d^\Sigma).
\end{equation}

To show that $\mu(\Lambda)$ is independent of $T$ we combine the equations
(\ref{egenekv}) to get
\begin{eqnarray}
\frac{6\mu(\Lambda)}{\mu_d^\Sigma}=3D&&(\Delta u+4\Delta d+\Delta s)
(1+T)+(4\Delta u-2\Delta d+4\Delta s)U\nonumber\\
=3D&&(T-1)((2A-B)\Delta u+(2C-2A)\Delta d+(B-2C)\Delta s).
\end{eqnarray}
Using (\ref{musp}) and (\ref{musm}) we obtain
\begin{equation}
\mu(\Lambda)=3D\frac{\mu(\Sigma^+)-\mu(\Sigma^-)}{6}
\left(2AR+
2C \left(1-R\right)-B\right),
\end{equation}
where
\begin{equation}
 R=3D\frac{\Delta u -\Delta d}{\Delta u-\Delta s}.
\end{equation}
Thus for  $\mu(\Lambda)$ to be independent of $T$ it is sufficient
that the ratio $R$ is independent of $T$.
To see that this is indeed so, we combine equations (\ref{egen1}),
(\ref{egen2}) to obtain
\begin{equation}
 R=3D\frac{\Delta u -\Delta d}{\Delta u-\Delta s}=3D \frac{B-C}{D-C},
\end{equation}
 which finally gives
\begin{equation}
\mu(\Lambda)=3D\frac{\mu(\Sigma^+)-\mu(\Sigma^-)}{6}
\left(\frac{2AB-2AC+2CD-BD-BC}{D-C}\right).
\end{equation}

A similar analysis can be made for $\mu(\Sigma \Lambda)$ giving
\begin{equation}
\mu(\Sigma\Lambda)=3D-\frac{\mu(\Sigma^+)-\mu(\Sigma^-)}{2\sqrt{3}}
\left(\frac{2B-C-D}{D-C}\right).
\end{equation}
The predictions of these moments are summarized in Table \ref{tabellen}
together with experimental data.
The value of $\mu(\Lambda)$ which we have found agrees very well with
experiments. However, our value of $\mu(\Sigma \Lambda)$ is somewhat low
compared to the experimental value $-1.61\pm0.08$. This value
is the average of two experiments which obtained the values
$-1.72_{-0.19}^{+0.17}$ \cite{dydak} and $-1.59\pm0.12$ \cite{petersen}
respectively.
We note that the errors in the individual measurements are quite large.
We therefore do not take this slight disagreement too seriously.

Since we cannot use further magnetic moment data as input
to determine the parameters
of the model, we proceed to discuss the weak axial-vector form factors.

\section{The weak axial-vector form factors}
=46or the weak decays we can express
the axial-vector form factors $g_A/g_V\equiv g_{A}$ in terms of
the spin polarizations as
\begin{mathletters} \label{ga}
\begin{eqnarray}
g_{A}^{np}=3D&&\Delta u-\Delta d, \label{ganp}\\
g_{A}^{\Lambda p}=3D&&\frac{1}{3}(2\Delta u-\Delta d-\Delta s),\\
g_{A}^{\Xi \Lambda}=3D&&\frac{1}{3}(\Delta u+\Delta d-2\Delta s),\\
g_{A}^{\Sigma n}=3D&&\Delta d-\Delta s. \label{gaSn}
\end{eqnarray}
\end{mathletters}

The expressions for the weak form factors above immediately leads to
two independent sum-rules.
They can be chosen to be
\begin{eqnarray}
(0.98\pm0.07)\quad g_{A}^{\Xi \Lambda} + g_{A}^{\Lambda p} =3D &&
g_{A}^{\Sigma n}+g_{A}^{np}\quad (1.06\pm0.08)\\
(1.51\pm0.05)\quad g_{A}^{\Xi \Lambda}+ g_{A}^{np}=3D&& 2g_{A}^{\Lambda p}
\quad (1.46\pm0.03).
 \end{eqnarray}
They are both satisfied within the experimental errors, indicating that the
model is correct within the error limits though there are in principle
still
some minor corrections to the expressions for the $g_A$'s due to
SU(3) symmetry breaking in the wave functions to be applied
\cite{donoghue}.

The experimental values for $g_A^{np}$ and $g_A^{\Xi\Lambda}$
are taken from the Review of Particle Properties (RPP) data table
\cite{data}, while the other two are not for the following reason.

 The value of $g_A^{\Sigma n}=3D-0.340\pm0.017$
listed in RPP is actually
$g_A-0.237 g_2$, which reduces to $g_A$ when $g_2=3D0$ in the axial-vector
current
\begin{equation}
A_\mu=3Dg_A\gamma_\mu\gamma_5-g_2\frac{i\sigma_{\mu\nu}q^\nu\gamma_5}{m_i+m_=
j}.
\end{equation}
The measurements and analysis made by Hsueh et al. \cite{hsueh}
give the values
\begin{eqnarray}
g_2=3D&&0.56\pm0.37,\\
g_A=3D&&-0.20\pm0.08.
\end{eqnarray}

A similar comment applies to $g_A^{\Lambda p}$. The value
in the RPP data table, $g_A^{\Lambda p}=3D0.718\pm0.015$,
 is given with the assumption that the weak magnetism
coupling $g_W=3D0.97$, but measurements by Dworkin et al. \cite{dworkin}
present the value $g_W=3D0.15\pm0.30$ which then gives $g_A=3D0.731\pm0.016$=
{}.

Needless to say all the equations (\ref{ga}) for the form factors
are independent of $T$. Hence also the sum-rules connecting them are
independent
of $T$.

The set of magnetic moment data and the axial-vector form factors can now
be connected to each other
by comparing the ratio $R$ calculated for both cases.
It is given by
\begin{eqnarray}
R&=3D&\frac{\Delta u -\Delta d}{\Delta u-\Delta s} =3D \frac{B-C}{D-C}  =3D
1.18 \pm 0.01, \\
R&=3D& \frac{2g_A^{np}}{g_A^{\Lambda p}+g_{A}^{\Xi \Lambda}+g_{A}^{\Sigma n}=
+
g_A^{np}}=3D 1.23 \pm 0.06.
\end{eqnarray}
The agreement is satisfactory, showing that the spin polarizations in the
magnetic moments and in the weak axial-vector form factors can be considered=
 to
be the same.

As input we choose $g_A^{np}$ in (\ref{ganp}) to be given by the
experimental
value $1.2573\pm0.0028$, as is generally done when analyzing spin
polarization data.
The other form factors are then functions of $g_A^{np}$ and $R$ and therefor=
e
do not supply any further information.

The predictions of these weak axial-vector form factors
are summarized in Table \ref{tabellen} together with experimental data.

The $g_A$'s agree well with the data. The dominating errors of
the $g_A$'s given in Table~\ref{tabellen}, come from the
neglected corrections mentioned above which are about 4\%.

\section{Symmetry breaking} \label{symmetry}

We now proceed to determine the symmetry breakings $T$ and $U$.
In order to be able to solve for $T$ we need one further input parameter,
which we have chosen as $\Delta \Sigma \equiv \Delta u+\Delta d+\Delta s$
obtained from measurements of the spin structure functions
\cite{emc,smc,e142}. Since $\Delta \Sigma$ is still not too well determined,
it is not clear which value to use. We therefore have to limit ourselves to
discuss the relation between $\Delta\Sigma$ and $T$ and are unable to give
any definite prediction for $T$.
$\Delta \Sigma$ as a function of $T$ is given by
\begin{equation}
\Delta \Sigma=3Dg_A^{np}\frac{D(B+C+D)-3BC}{(B-C)(3-D\Delta T)}\Delta T,
\end{equation}
where
\begin{equation}
\Delta T=3D\frac{T-1}{T+1}.
\end{equation}
 In Figure \ref{fig} we show $\Delta\Sigma$ as a function of $T$
with the error band coming from the other input data.  We see that to
have a small isospin breaking ($\leq 5\%$) we cannot have $\Delta \Sigma$
larger than
about 0.20. Demanding isospin symmetry gives $\Delta\Sigma=3D0.08\pm 0.05$.

Analyses of data from deep inelastic scattering measurements
have recently been made presenting different values for $\Delta\Sigma$.
Ellis and Karliner \cite{ellis} have found the value $\Delta \Sigma=3D0.27
\pm0.11$. Close and Roberts \cite{close} have made a similar analysis and
found $\Delta \Sigma=3D0.38\pm0.48$. The major differences between these
analyses is the treatment of the higher twist contributions. Ellis and
Karliner
have used theoretical predictions \cite{balitsky}. Alternatively
Close and Roberts have treated these corrections as free parameters and
determined their
values by making the data from EMC, SMC and E142 measurements consistent
with each other.

{}From Figure \ref{fig} we see that the analysis of Ellis and Karliner gives
$T=3D\mu_u/\mu_d=3D-1.79^{+0.02}_{-0.14}$. This is in agreement with
lattice calculations which give $\mu_u/\mu_d=3D-1.76\pm0.15$
\cite{leinweber}.
Similar results have also been found in the chiral bag model \cite{krivo}
and in
analysis of magnetic transitions moments of vector mesons \cite{geffen}.

The value given by Close and Roberts, on the other hand, admits a large
range of
$T$-values that includes $T =3D -2$, i.e. no isospin symmetry breaking.

A more recent analysis from the SMC collaboration \cite{SMC} gives the
value $\Delta \Sigma=3D0.24\pm0.23$ which also includes the value $T=3D-2$ i=
n
its range of
admissible $T$-values.

We conclude that a large value of $\Delta\Sigma$ of the order of $0.75$
as a solution of the so called ``spin crisis", compatible with the
analysis of Close and Roberts, forces the isospin symmetry breaking to be
rather large;
the spin crisis is resolved at the cost of an ``isospin crisis".
A small value of $\Delta \Sigma$ on the other hand, cannot give any
prediction at all
for the size of the isospin symmetry breaking, and is compatible with no
isospin symmetry breaking; see Figure \ref{fig}.

The value of the hypercharge symmetry breaking $U=3D\mu_s/\mu_d$ is in the
range $0.67$ to $0.70$ for $T$ in the range $-2$ to $-1.75$. This
can for example be compared with $m_{u,d}/m_s=3D0.70\pm0.07$ obtained from a=
n
analysis of radiative $K^*$ decays \cite{bramon}, though the effective
quark masses might not be the same in mesons and baryons.

\section{Conclusions}

We have studied baryon magnetic moments, weak axial-vector
form factors and quark spin polarizations in a model that admits flavor
symmetry breaking and quark magnetic moments varying from isomultiplet to
isomultiplet.
Several physical quantities, such as the weak axial-vector form factors,
and the magnetic moment of $\Lambda$, are independent of this symmetry
breaking, whereas the spin polarization of the quarks do depend on the symme=
try
breaking. The model reproduces weak axial-vector form factors and
$\mu(\Lambda)$ very
well and $\mu(\Sigma\Lambda)$ rather well.

The analysis shows that already a spin polarization of the nucleon of
the order of $0.35$ or greater requires an isospin
symmetry breaking that is of the order of $10\%$. A small symmetry breaking
($\leq 5\%$) is only compatible with a small value of the nucleon spin
polarization.

\section*{Acknowledgement}

One of us (HS) would like to thank G. Karl for inspiring discussions.

\begin{figure}
\caption{$\Delta\Sigma$ as a function of the isospin symmetry breaking $T$.
The error
band comes from the errors in the magnetic moment input data.}
\label{fig}
\end{figure}

\mediumtext

\begin{table}
\caption{Prediction of physical quantities independent of the symmetry
breaking
compared with experimental data.
 Magnetic moments are given in units of $\mu_N$.
Data is taken from \protect\cite{data}, unless otherwise stated.}
\label{tabellen}
\begin{tabular}{ldd}
&Theory&Experiment\\
\tableline
$\mu(\Lambda)$&$-$0.63${}\pm{}$0.03&$-$0.613${}\pm{}$0.004\\
$\mu(\Sigma\Lambda)$&$-$1.41${}\pm{}$0.01&$-$1.61${}\pm{}$0.08\\
$g_{A}^{np}$&1.2573${}\pm{}$0.0028 (input)&1.2573${}\pm{}$0.0028\\
$g_{A}^{\Lambda p}$&0.78${}\pm{}$0.04
&0.731${}\pm{}$0.016\tablenotemark[1]\\
$g_{A}^{\Xi\Lambda}$&0.29${}\pm{}$0.02&0.25${}\pm{}$0.05\\
$g_{A}^{\Sigma n}$&$-$0.19${}\pm{}$0.02
&$-$0.20${}\pm{}$0.08\tablenotemark[2]
\end{tabular}
\tablenotetext[1]{From Dworkin et al. \cite{dworkin}}
\tablenotetext[2]{From Hsueh et al. \cite{hsueh}}
\end{table}

\narrowtext

\end{document}